\documentclass[aps,twocolumn,prc,epsfig,showpacs]{revtex4}
\usepackage{graphicx}
\usepackage{epsfig}
\begin{document}
\title{Coherent dimer formation near Feshbach resonances in Bose-Einstein 
condensates}

\author{Victo S. Filho$^{1}$, Lauro Tomio$^{1}$, Arnaldo Gammal$^{2}$ and
T. Frederico$^{3}$}
\affiliation{$^{1}$Instituto de F\'{\i}sica Te\'{o}rica,
Universidade Estadual Paulista, 01405-900, S\~{a}o Paulo, Brazil\\
$^{2}$Instituto de F\'\i sica, Universidade de S\~ao Paulo,
05315-970, S\~ao Paulo, Brazil\\
$^3$Dep. de F\'\i sica, Inst. Tecnol\'{o}gico da Aeron\'{a}utica,
CTA, 12228-900, S. Jos\'{e} dos Campos, Brazil}
\date{\today}

\begin{abstract}
The results of a recent experiment with $^{85}$Rb Bose-Einstein
condensates are analyzed within the mean-field approximation 
including dissipation due to three-body recombination. 
The intensity of the dissipative term is chosen from the three-body 
theory for large positive scattering lengths.
The remaining number of condensed atoms in the experiment, obtained 
with applied magnetic field pulses, were used to adjust the intensity 
of the dissipative term. We found that the three-body recombination 
parameter depends on the pulse rise time; 
i.e., for longer rise times the values found
become consistent with the three-body theory, while for shorter
pulses this coefficient is found to be much larger. We interpret
this finding as an indication of a coherent formation of dimers.
%%% 03.75.-b Quantum condensation phenomena
%%% 21.45.+v Few body systems
%%% 36.40.-c Atomic and Molecular Clusters
%%% 34.10.+x General Theories and models of atomic...
%%% \vskip2pc]
\end{abstract}

\maketitle

Bose-Einstein condensation of dilute atomic gases has been observed for
the first time in $^{87}$Rb atoms~\cite{Cornell}. Since then, it has
been condensed atoms of $^{23}$Na~\cite{Ketterle},
$^{7}$Li~\cite{Hulet}, $^1$H~\cite{Fried},
metastable $^4$He~\cite{Cohen}, $^{85}$Rb~\cite{Cornish},
$^{41}$K~\cite{Modugno} and, more recently, $^{133}$Cs~\cite{Tinoweber}.
In Ref.~\cite{Tiesinga}, it was theoretically predicted that Feshbach
resonances could vary the scattering length of atoms in systems of 
dilute alkali gases over a wide range of values. 
A Feshbach resonance occurs when the quasimolecular bound state
energy is tuned to the energy of two colliding atoms, applying an
external magnetic field to the system. The phenomenon was first
realized in a Bose-Einstein condensate in Ref.~\cite{Inouye}.
This opened the possibility of exploring new regimes of BEC, allowing
changes of the two-body scattering length from negative to
positive and also from zero to infinity.
This technique was used for the condensation and collapse control of
$^{85}$Rb atoms in the hyperfine state
($F = $ 2, $m_F=-2$)~\cite{Cornish,Roberts,Donley}.
It was also demonstrated in Ref.~\cite{Stenger} that strongly enhanced
inelastic three-body collisions occurs near Feshbach resonances.
In a more recent experiment, in Ref.~\cite{Claussen}, it was explored the
region of very large scattering lengths (up to $\sim 4000$ Bohr radius).

The scattering length $a$ has been observed to vary as a function
of the magnetic field $B$, according to the theoretical
prediction~\cite{Moerdijk}, as
\begin{equation}
a=a_{b} \times \left(1-\frac{\Delta}{B-B_r}\right)\,\,\,,
\label{eq1}
\end{equation}
where $a$ is the scattering length, $a_b$ is the background scattering
length, $B_r$ is the resonance magnetic field of $^{85}$Rb, and
$\Delta$ is the resonance width.

In the case of $^{85}$Rb, one has resonance width $\Delta \cong $
11.0 G, resonance field $B_{r} \cong $ 154.9 G and background
scattering length  $a_b \cong -450 a_0$~\cite{Claussen}, where
$a_0$ is the Bohr radius. So, given the experimental functional
dependence $B=B(t)$,  one can determine $a=a(t)$ and,
consequently, the dynamics of the system in such physical
conditions. Further, mainly in strong interaction regime, a
Bose-Einstein condensate shows inelastic loss processes that cause
its depletion. The dominant process of losses has been verified to
be the three-body recombination \cite{Claussen,Burt}, with a time
dependence concerning a simple constant rate equation. All earlier
observations in BEC experiments were consistent with a description
of mean field including such a loss process. But, in experiments
with $^{85}$Rb realized in the strong interaction regime
\cite{Claussen}, this picture indicates model breakdown.
Bose-Einstein condensates initially stable were submitted to
magnetic field pulses carefully controled in the vicinity of
$^{85}$Rb Feshbach resonance, aiming to test the strongly
interacting regime for diluteness parameter $\chi= n a^3$ varying
from $\chi=$ 0.01 to $\chi=$ 0.5. The loss of atoms from BEC
occurred in impressively short time scales (up to two hundreds of
$\mu$s) and disagrees with previous theoretical predictions
\cite{Cornish}. Such experiments reveal higher loss of atoms in
shorter magnetic field pulses applied on BEC and, previously, 
one knew that as longer is the time spent near a
Feshbach resonance as higher is the loss of atoms from BEC
~\cite{Cornish} (consistent with a mean field approach
where the inelastic loss term has a constant dissipative rate). 
According to Ref.~\cite{Claussen}, the results could indicate the
existence of a new physics, that cannot be described by
the Gross-Pitaevskii (GP) formalism.
Motivated by this observed discrepancy, we investigate
the dynamics of $^{85}$Rb Bose-Einstein condensates submitted to such
external conditions and time scales, when we vary the {\it s}-wave
two-body scattering length in the region of strong interaction. 
So, we consider a generalized mean field approach that includes
the time dependence for both mean field coupling and three-body
recombination parameter. Previous results considered a mean field
approach with constant mean rate for the three-body recombination
and constant value for the mean field coupling~\cite{Kagan,Saito}. 
In the present work, our first task is to reproduce the
experimental data with time dependent parameters
or at least verify possible limitations of the time dependent
mean field approach. Next, we consider the magnitude of the
recombination rate as described in
literature~\cite{Fedichev,Nielsen1,Esry,Bedaque,Tomio,Nielsen2,aform1,aform2}, 
but with time functional dependence.

For describing a BEC in the framework of the mean field approximation,
we remind that such an approach is valid for very dilute systems when the 
average inter particle distances $d$ are much larger than $|a|$
and the particle wavelengths are much larger than $d$~\cite{Yukalov,Dalfovo}.
Besides, it is important to pay attention to the time scales present in the
system: the physical conditions must not change fast enough in order to
allow the replacement of a true interatomic potential by the contact
interaction. It is possible in principle for the rate of change to be larger 
than $\hbar/ma^2$ for extremely very short changes in the interactions
~\cite{Legget}, but it is reasonable to assume valid this time dependent approach 
at least for longer pulses.  

We begin our description from an effective Lagrangian of the nonconservative
system  as in \cite{Abdullaev}, in which one describes the dynamics of a
trapped Bose-Einstein condensate in spherical symmetry with such a GPE
generalization, in which one also considered losses from BEC by
three-body recombination. This Lagrangian leads to the following
equation of the system:

\begin{eqnarray}
i \hbar\frac{\partial\Psi}{\partial t}=\left[
-\frac{\hbar^2}{2m}\nabla^2
+\frac{m\omega^2r^2}{2}+U_0|\Psi|^2
-{\rm i} \hbar \frac{K_3}{4} |\Psi|^{4}
\right]\Psi,
\label{3}
\end{eqnarray}
where
$\omega$ is the geometric mean trapping frequency,
$U_0\equiv U_0(t)=4\pi\hbar^2a(t)/m$ and $K_3\equiv K_3(t)$ is the
recombination loss parameter. The wave-function $\Psi=\Psi(\vec{r},t)$
is normalized to the number of atoms $N$.
The three-body recombination rate $K_3$ is here introduced
for describing atomic losses from the condensate when three atoms
scatter to form a molecular bound state (dimer) and a third atom;
so, the kinetic energy of the final state particles allows them to
escape from the trap. Other nonconservative processes as 
amplification from thermal cloud and dipolar relaxation are
neglected, since the latter has a much smaller effect
than three-body recombination \cite{Claussen,Burt} and in JILA
experiments \cite{Claussen} the thermal cloud is negligible (only
1000 atoms in a sample of 17500).

A theoretical prediction of $K_3$ is a hard task since
it is sensitive to the detailed behavior of the interaction
potential \cite{aform1,aform2}. However, such a calculation becomes
simpler if we consider that $a$ is the only important length scale
(reasonable in the weakly bound {\it s}-wave state limit) and
this has been considered in many
works~\cite{Fedichev,Nielsen1,Esry,Bedaque,Tomio,Nielsen2,aform1,aform2}.
Following Ref.~\cite{Fedichev}, the recombination rate is written as 
\begin{equation}
K_3(t) \cong \kappa\frac{\hbar}{m}[a(t)]^4\,\,\,,
\label{eq100d}
\end{equation}
where $\kappa$
should correspond to the universal value $\kappa = $ 3.9. But in
Ref. \cite{Nielsen1} and in Ref. \cite{Esry}, it was found $0 \le
\kappa \le 65$ and in Ref. \cite{Bedaque}, $0 \le \kappa \le
67.9$, that is, $\kappa$ is not universal (in general $K_3$
depends on a three-atom scale\cite{Tomio}).
%%%%%%%%%%%%%%%%%%%%%%%%%%%%%%%%%%%%%%%%%%%%%%%%%%%%%%%%%%%%%%%%%%%%%%%%%
%%%Fig. 1
% \vskip -.6cm
\begin{figure}
 \setlength{\epsfxsize}{0.8\hsize} \centerline{\epsfbox{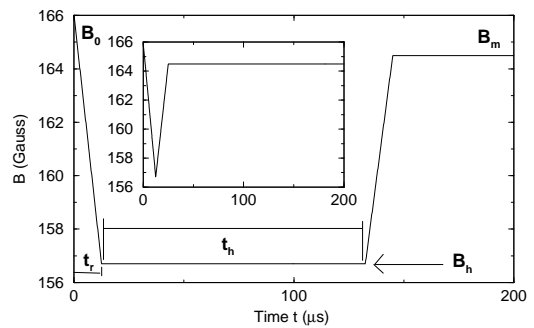}}
 \caption[dummy0]
{Triangular (insertion) and trapezoidal shapes of magnetic field
 pulse (Gauss) applied to $^{85}$Rb BEC
 as function of pulse time ($\mu $s). The rise time $t_r$ of the pulse
amounts to 12.5 $\mu$s and the hold field is 156.7 G (scattering length
 $a\cong $ 2300a$_0$). The hold time of trapezoidal pulse is 120 $\mu$s.}
 \label{campo}
 \end{figure}
%%%%%%%%%%%%%%%%%%%%%%%%%%%%%%%%%%%%%%%%%%%%%%%%%%%%%%%%%%%%%%%%%%%%%%%%%%
In Fig.~\ref{campo}, we schematically give two
characteristic pulses employed
in JILA, in which $B_0$, $B_h$, $B_m$, $t_r$ and $t_h$,
correspond to the initial field, the hold field, the final field, the rise
time and the hold time of the pulse, respectively.
For describing the dynamics of condensates subjected to (\ref{eq1}),
we consider in our calculations the same experimental parameters and
conditions used in \cite{Claussen}: as $a$ is known to be a
function of the magnetic field $B$ by means of (\ref{eq1}), we
only use time dependent shapes of experimental magnetic field
pulses employed in JILA (triangular and trapezoidal pulses). The
hold field is $B_h=$ 156.7 G, and the end field in which one
measures the remaining number of particles $N_r$ of the system is
$B_m=$ 164.5 G at $t$ = 700 $\mu$s. Further, we put initial field
$B_0 \cong 166$ G, corresponding to a harmonic oscillator state of
the system ($a\sim 0$), applied to an initial sample of
$N_0$=16500 condensed atoms of $^{85}$Rb. Further, in our approach
we have used spherical symmetry with mean geometric frequency
$\omega=(\omega_r ^2\omega_z)^{1/3}$, for simulating the
cylindrical geometry of JILA (radial: $\omega_r = 2\pi\times 17.5$
Hz and axial: $\omega_z = 2\pi\times 6.8$ Hz). Our time dependent
calculations started with a Gaussian shape wave-function which we numerically 
evolve by means of Eq.~(\ref{3}), using the Crank-Nicolson
algorithm, as in Refs. \cite{caos,pre}. We analyzed the loss of
condensed atoms like in \cite{Claussen}, by considering hold times
from  $t_{hold}=$ 0 (triangular) or units of $\mu$s (shorter
trapezoidal pulses) to nearly hundreds of $\mu$s (longer
trapezoidal pulses).
 The behavior of the scattering length, as a function of total
time of the pulse, in the region of strong interaction atom-atom,
follows similarly the behavior of the employed field pulses given
in Fig.~\ref{campo}. In triangular pulses, there is a sharp peak in the 
resonance region; and a plateau with maximum value for $a$ and $\xi$ 
(minimum value of the field), when $B=B_h$ is kept constant during the time 
$t_h$. 
%%%%%%%%%%%%%%%%%%%%%%%%%%%%%%%%%%%%%%%%%%%%%%%%%%%%%%%%%%%%%%%%%%%%%%
%%%Fig. 2
 \begin{figure}
 \setlength{\epsfxsize}{0.78\hsize}\centerline{\epsfbox{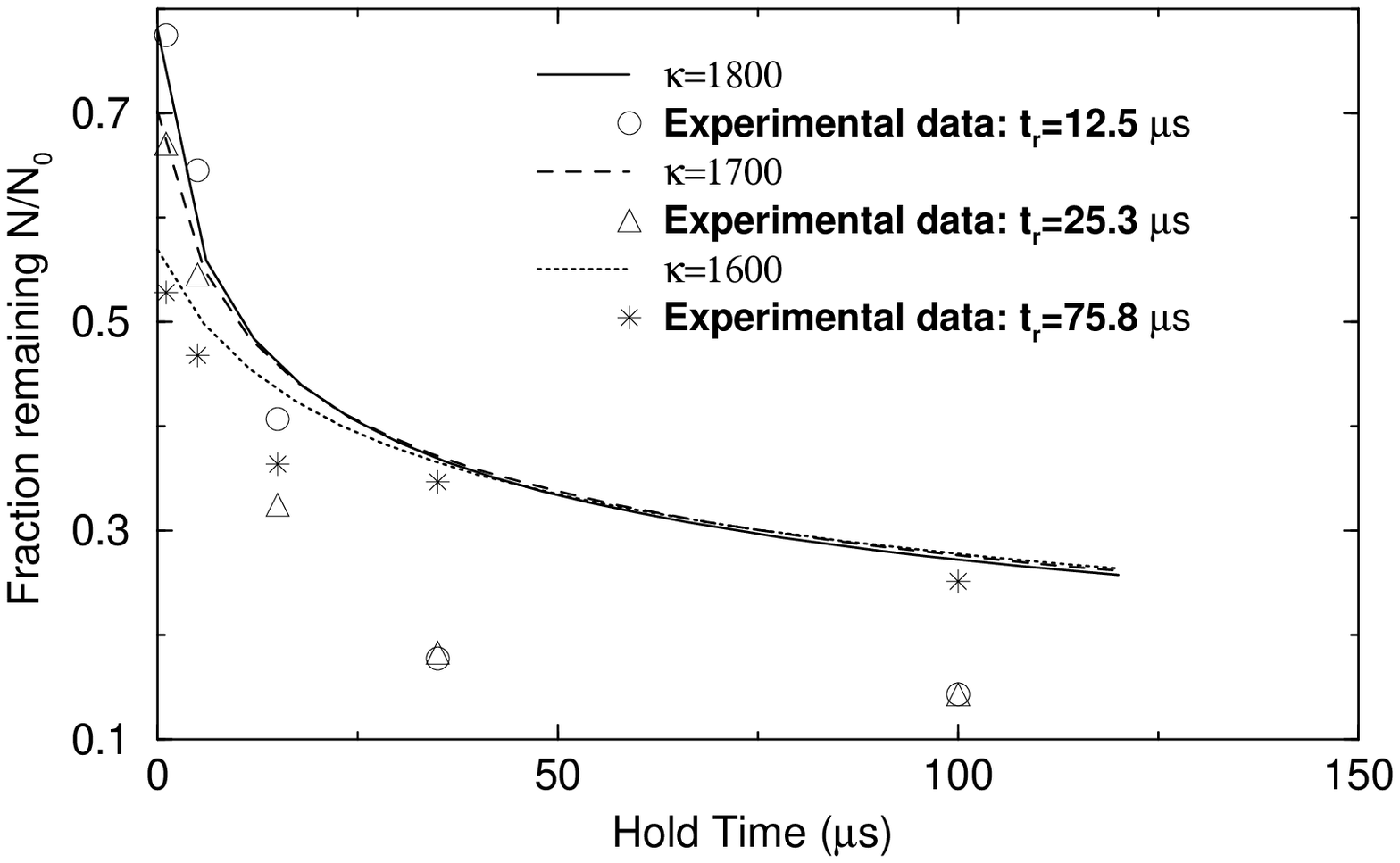}}
 \setlength{\epsfxsize}{0.78\hsize}\centerline{\epsfbox{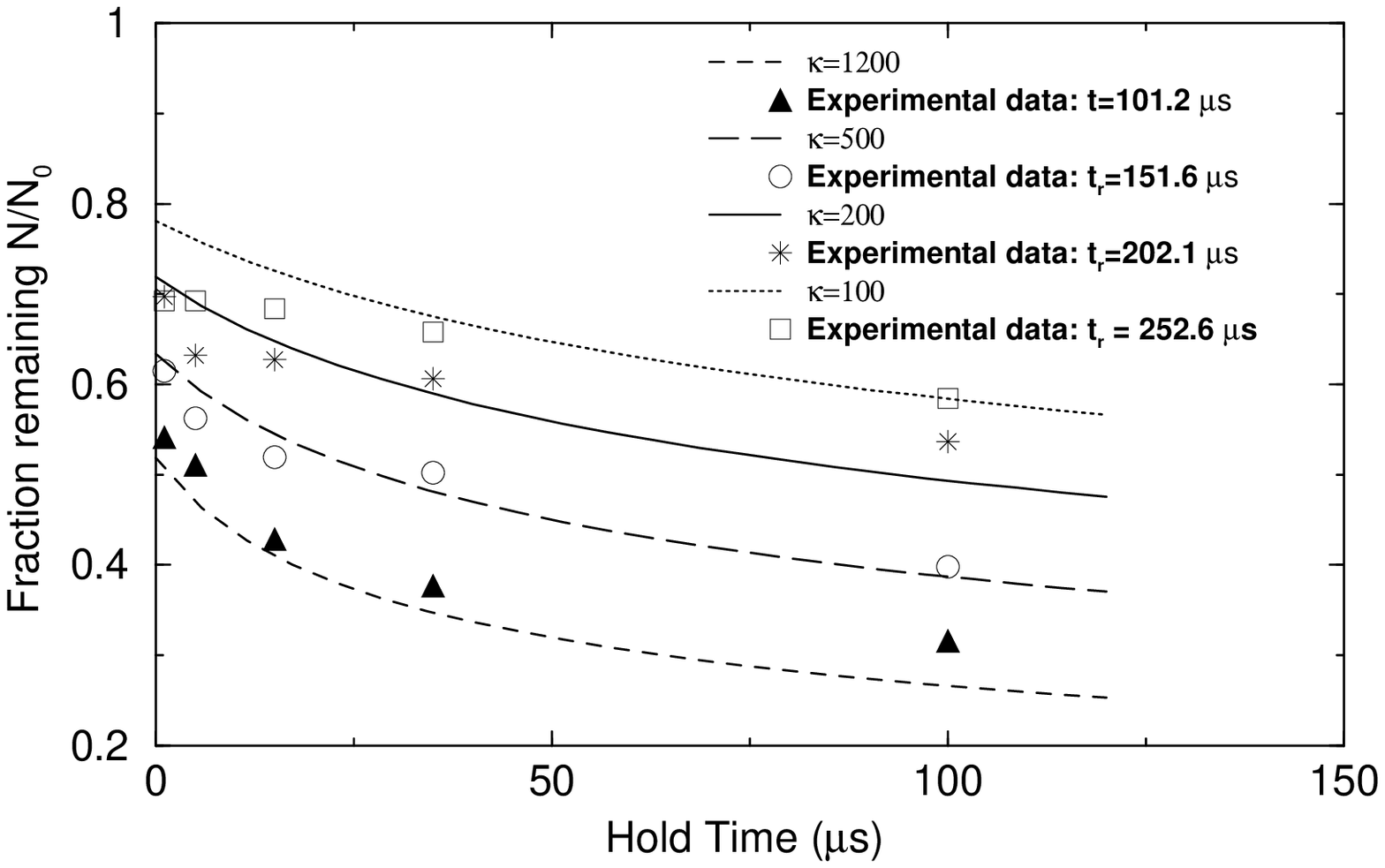}}
 \caption[dummy0]{Remaining fraction of atoms in $^{85}$Rb condensate
  versus hold time for some rise times (indicated inside frames).
  For upper frame, $\kappa \cong $ 1800 and it decreases up to
  $\kappa \cong$ 100 (lower). The initial number of particles is 16500.
  $N_r$ is calculated as if it was measured at $\tau$ = 700 $\mu$s
  ($B_m=$164.5 G).}
 \label{fracaorest}
 \end{figure}
%%%%%%%%%%%%%%%%%%%%%%%%%%%%%%%%%%%%%%%%%%%%%%%%%%%%%%%%%%%%%%%%%%%%%%

In the upper frame of Fig.~\ref{fracaorest}, we used symmetric
rise and fall times ($t_r=12.5 \mu$s, $t_r\cong 25 \mu$s and 
$t_r\cong 75 \mu$s), to determine the remaining fraction of atoms $N_r/N_0$ in
the $^{85}$Rb BEC as function of the hold time, by adjusting our curve 
with the first point of Fig.~\ref{fracaorest} in Ref.~\cite{Claussen}. 
For $t_r=12.5\mu$s, we found $\kappa\cong 1800$, very far away of 
the values described earlier in the literature
~\cite{Fedichev,Nielsen1,Esry,Bedaque,Tomio,Nielsen2,aform1,aform2}. 
The results show the same exponential decay and good concordance with
experimental data, mainly for short $t_h$ (circles in upper
frame of Fig.~\ref{fracaorest}). However, as one can realize, for
longer hold times, experimental data point out a higher
dissipation when compared with our simulations. For other short 
rise times in this frame or longer rise times (lower frame), 
we have similar behavior, but we have to decrease $\kappa$ for better 
adjusting to experimental data. 
So, the comparison with experiments show that $\kappa$ depends
significantly on values of $t_r$ and $t_h$. As we know
\cite{Stoof}, the mean field approach should make more adequately
if we were in a slower process. So, we also tried to verify if our
results would give a lower value of $\kappa$ (inside interval
described in literature) if we calibrated our calculation with the
last point of the longest pulse of JILA \cite{Claussen} ($t_r
\cong $ 252.6 $\mu$s). Really, we found $\kappa\cong 100$ for this
case (closer to values described in literature
\cite{Fedichev,Nielsen1,Esry,Bedaque,Tomio,Nielsen2,aform1,aform2}) and the
results reproduce the experimental data for longer  rise time but
they do not make very well for shorter rise time, as we can
observe in lower frame of Fig.~\ref{fracaorest}.
The very large value of $\kappa$ leads us to conclude that there
is a coherent formation of dimers, that occurs up to nearly $t_r
\sim $ 100$\mu s$, as one can realize observing the lower frame of
Fig.~\ref{fracaorest} for 
%%%????longer?????
shorter rise times.
%%%( NAO ENTENDI: where there
%%%is a better agreement between experimental data and simulations
%%%for all hold times).
It is physically sensible that, for shorter pulses, the presence
of coherence in the formation of dimers would be more plausible
than for longer pulses.
So, the present calculation included the variation of $\kappa$ 
with the time parameters of the pulse and the amplification of 
the three-body recombination rate can be associated with the 
coherent formation of dimers in the inelastic collisions.  
Together with the significant loss
from the coherent formation of dimers, one should also observe a
burst of atoms carrying the excess of energy, which for $^{85}$Rb
with the maximum value of $a=4000 a_0$, would be above 70nK.
Indeed, in the JILA experiment, it was seen a significant number
loss from the BEC for pulses lasting only few tens of
microseconds, which were accompanied by a {\it burst} of few
thousand relatively hot ($\sim $ 150nK) atoms that remained in the
trap \cite{Claussen}. In our description, for each hot atom
one dimer is also formed. Therefore, the burst of atoms should be
accompanied by a burst of weakly bound dimers. Using the observed
temperature and momentum conservation of the recombination
process, we predict that the dimers are also relatively hot
($\sim$ 75nK).

%Lauro:
We note that, if we consider a constant value of $\kappa$, in all
the cases (for any choices of $t_h$) we obtain a decreasing
behavior of $N_r$, as we increase the rise time $t_r$. This is in
contrast with the experimental results. So, we consider to adjust
the values of $\kappa$ that approximately better describe the
experimental data for each given rise time $t_r$; i.e., $\kappa$
is taken as a function of the rise time ($\kappa=\kappa(t_r)$).
Our results are shown in Table I.
There is an obvious uncertainty in the given numbers of Table I,
that are related to our approximate theoretical fitting and
experimental data fluctuations. However, based on such
results, the behavior of $\kappa(t_r)$ can be approximately
described by a linear or exponential decreasing function.
Here we consider the following simple functional time-rise
dependences of $\kappa$:
\begin{eqnarray}
\kappa(t_r)&=&2300\exp{(-0.01\times\omega t_r)}\;\;\;\textnormal{or}
\label{adj1}\\
\kappa(t_r)&=&1900-7\omega t_r\;.\label{adj2}
\end{eqnarray}

\begin{figure}
 \setlength{\epsfxsize}{0.78\hsize}\centerline{\epsfbox{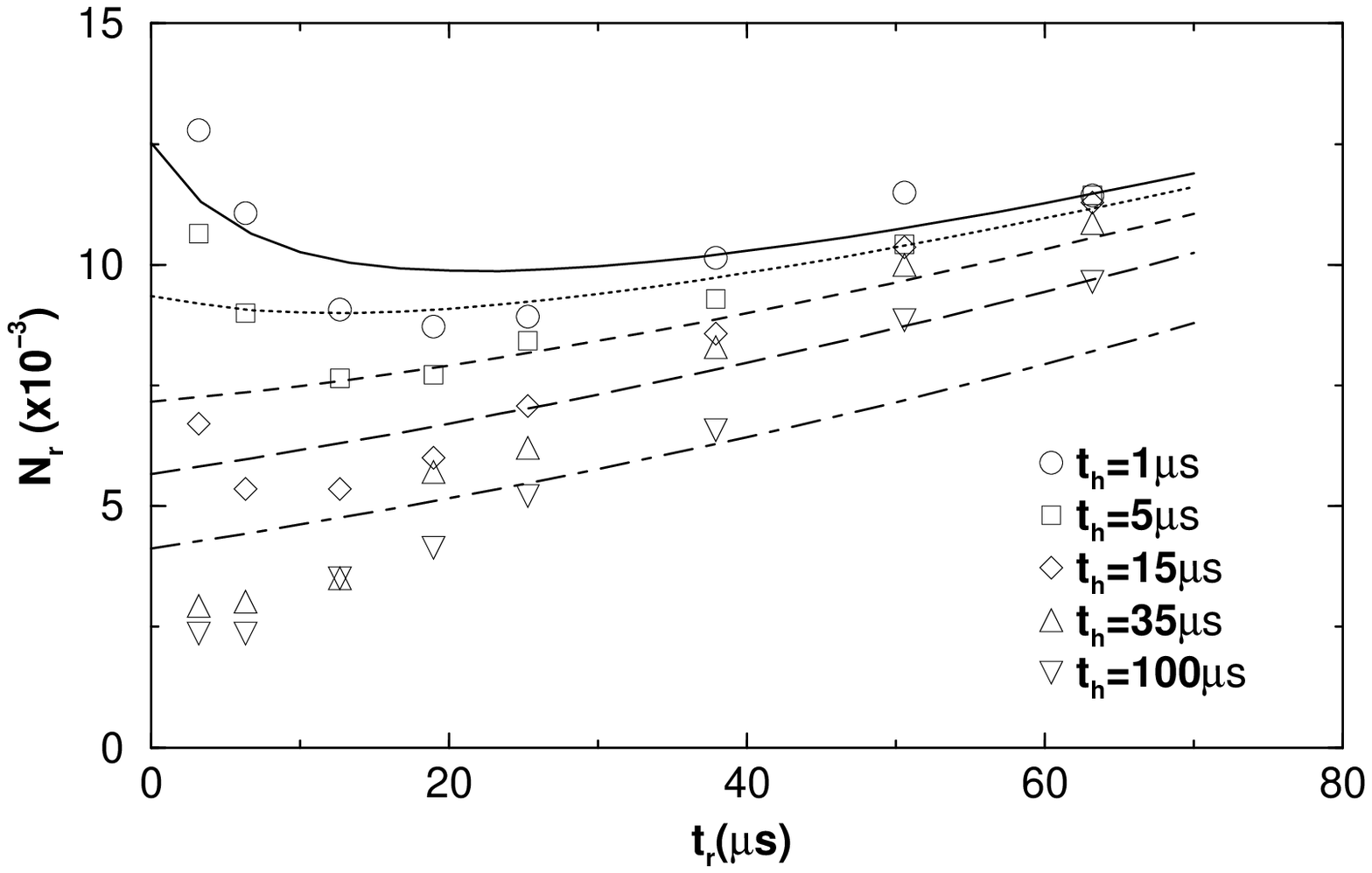}}
 \setlength{\epsfxsize}{0.78\hsize}\centerline{\epsfbox{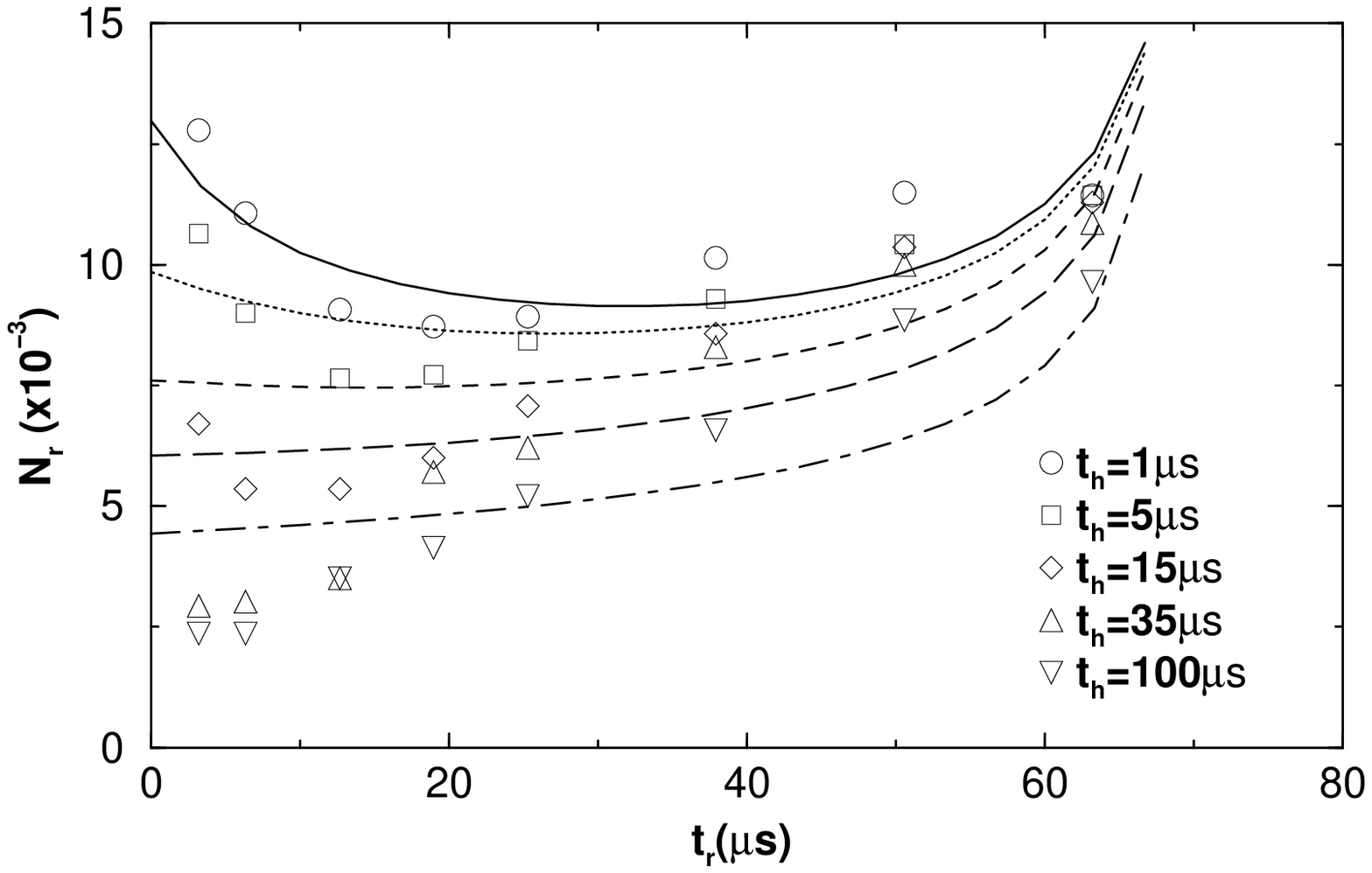}}
 \caption[dummy0]{Remaining number of atoms in
 $^{85}$Rb BEC versus the scaled rise time (factor 1/4)  of the
applied magnetic pulse, for some hold times, with hold field
156.7 G (2300$a_0$) and initial number $N_0$=16500.
In the upper frame, we consider $\kappa$ given by the exponential
dependence of Eq.~(\ref{adj1}); in the lower frame, we
consider $\kappa$ given by Eq.~(\ref{adj2}).}
 \label{numrest1}
 \end{figure}

\begin{table}
\caption{Numerical values of three-body recombination coefficient $\kappa$
as function of the rise time $t_r$ of the magnetic field pulses
applied to the $^{85}$Rb BEC.}
\begin{tabular}{cc|cc}
\hline\hline
 $t_r$ ($\mu$s) & $\kappa$ & $t_r$ ($\mu$s) & $\kappa$\\
 (shorter) & & (longer) &  \\
\hline
12.5 & 1800 & 151.6 & 500 \\
25.3 & 1700 & 202.1 & 200  \\
75.8 & 1600 & 252.6 & 100  \\
\hline\hline
\end{tabular}
\end{table}
With such decreasing functional time-rise dependence of
$\kappa$, we have verified, as shown in Fig.~\ref{numrest1}, that
we have the qualitative behavior observed in the experimental
results given in Ref.~\cite{Claussen} of the remaining number of
atoms $N_r$ versus $t_r$. We have considered several values of the
hold time $t_h$. For very small hold time, we can also
verify the same experimental results that presents a minimum of
$N_r$ as a function of $t_r$. With a linear functional $t_r$
dependence of $\kappa$, the same qualitative behavior can also be
reproduced; which differs quantitatively from the exponential
behavior that we show. So, we conclude that the time dependent
mean field approach can describe all the experimental data, if the
three-body recombination coefficient depends on the rise time in
this short time scale. Such dependence on the rise time can be
explored, once the uncertainties in the experimental results are
reduced and by considering a better fitting of data (improving the
values given in Table I). In our interpretation, the higher
values of $\kappa$ for smaller values of $t_r$ are indicating the
coherent formation of another species (dimers) in the condensate.
%%% TALVEZ FAZER UMA ANALOGIA COM O LASER EFFECT NESSE PONTO ???
%%%%%%%%%%%%%%%%%%%%%%%%%%%%%%%%%%%%%%%%%%%%%%%%%%%%%%%%%%%%%%%%%%%%%%
%%%Fig. 3
% \vskip -.5cm
%%%%%%%%%%%%%%%%%%%%%%%%%%%%%%%%%%%%%%%%%%%%%%%%%%%%%%%%%%%%%%%%%%%%%%%%%%
In summary,  we report in this work indications based on our
calculations that a recent experiment realized in
JILA~\cite{Claussen} is evidencing the coherent formation of
dimers from inelastic collisions in $^{85}$Rb Bose-Einstein
condensates (BEC). We have solved numerically the nonconservative
Gross-Pitaevskii equation in spherical symmetry, for condensed
systems with very large repulsive two-body interaction, varying in
time, due to application of magnetic field pulses, according to
Eq.~(\ref{eq1}). According to the theoretical predictions of
three-body recombination rates, we used a dissipation parameter 
proportional to the quartic power of the
scattering length $a(t)$ and the observed experimental pulse
shapes to calculate the time evolution of the remaining number of
atoms in the condensate. We studied this observable as a function
of the hold and rise times. The experimental results of $N_r$
versus $t_h$ can be described in the mean-field approach only with
very large $\kappa$, when the rise time is small. This indicates a
coherent formation of dimers in BEC. For longer pulses, when the
coherent dimer formation tends to disappear, we found that
$\kappa$ approaches the maximum value given by the theoretical
predictions for large scattering lengths, $\kappa\sim
70$~\cite{Nielsen1,Bedaque}. We parameterize this surprising
behavior of the three-body recombination rate considering
$\kappa=\kappa(t_r)$ and so it was possible to describe the property of  
a lower dissipation for longer pulses. 
Finally, it is natural to see a burst of relatively hot atoms and 
dimers (carrying the excess of dimer binding energy) accompanying a 
significant loss from the condensate for short pulses when the 
coherent dimer formation occurs.

We thank partial support from Funda\c c\~ao de Amparo \`a Pesquisa do
Estado de S\~ao Paulo (FAPESP) and Conselho Nacional de Desenvolvimento
Cient\'\i fico e Tecnol\'ogico (CNPq) of Brazil.

\end{document}